\documentclass[twocolumn,prb,showpacs,amsmath,amssymb]{revtex4}
\usepackage{graphicx}
\usepackage{dcolumn}
\usepackage{bm}

\begin{document}
\draft
\title{Quasi-Particle Dynamics in Superconducting Aluminum}
\author{Katrin Steinberg, Marc Scheffler, and Martin Dressel}
\address{1.~Physikalisches Institut, Universit\"at Stuttgart,
Pfaffenwaldring 57, D-70550 Stuttgart, Germany}
\date{Received  \today}
\begin{abstract}
The response of superconducting aluminum to electromagnetic
radiation is investigated in a broad frequency (45~MHz to 40~GHz)
and temperature range ($T>T_c/2$), by measuring the complex
conductivity. While the imaginary part probes the superfluid
density (Cooper-pairs), the real part monitors the opening of the
superconducting energy gap and -- most important here --  the
zero-frequency quasi-particle response. Here we observe the full
temperature and frequency dependence of the coherence peak.
Varying the mean free path gives insight into the dynamics,
scattering and coherence effects of the quasi-particles in the
superconducting state.

\end{abstract}

\maketitle

\section{Introduction}
\begin{figure}[h]
\includegraphics[width=60mm]{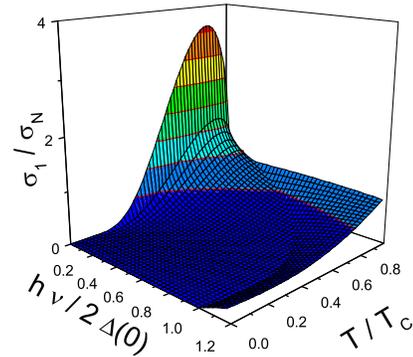}
\caption{\label{fig:BCS}(color online) Frequency and temperature
dependence of the real part of the conductivity
$\sigma_1/\sigma_n$ calculated according the BCS theory
\protect\cite{Mattis58,Zimmermann91} with the ratio of the coherence length to
the mean free path $\pi\xi(0)/2\ell =10$. The pronounced maximum for low frequencies at a temperature slightly below $T_c$ is the coherence peak.}
\end{figure}

Microwave and optical experiments have played a prominent role in
the elucidation of the superconducting state for more than fifty
years, because they provide information on the single-particle
excitations as well as on the response of the Cooper
pairs.\cite{Tinkham96,DresselGruner02}  The electrodynamic
properties of superconductors can be calculated on the basis of
the BCS theory,\cite{BCS} as first worked out by Mattis and
Bardeen;\cite{Mattis58} and there exist precise predictions on the
temperature and frequency dependence of the complex conductivity.
One of the hallmarks of the superconducting state is the opening
of an energy gap $\Delta$ in the density of states right below the
critical temperature $T_c$; both are related by mean-field theory:
$2\Delta = 3.53 k_BT_c$. In the 50s and 60s of the last century,
enormous efforts were undertaken to explore the microwave and THz
absorption of superconductors, like Al, In, or Sn, in order to
obtain the conductivity in the region of the superconducting
gap.\cite{Richards60} Much less is known about the response of the
quasi-particles in the superconducting state where, at energies
below the energy gap, a maximum in the electromagnetic absorption
-- the so-called coherence peak -- is expected for temperatures
slightly below the superconducting transition
(Fig.~\ref{fig:BCS}).
The analog feature in nuclear magnetic resonance, i.e.\ the
presence of a maximum in the nuclear-spin relaxation-rate of
aluminum below $T_c$,\cite{Hebel57} was one of the crucial
experiments for the quick success of the BCS theory. By now, this
Hebel-Slichter peak has been found in a large number of
superconductors. In the case of the electrodynamic response, the
experimental confirmation of the coherence peak turned out to be
much more difficult \cite{Waldram64} and it was only in the 1990s, that a maximum
in the temperature dependent conductivity was observed by studying
Pb and Nb in a microwave cavity at 60~GHz.\cite{Holczer91,Klein94} The complete mapping of the conductivity
coherence peak, in particular its frequency dependence, has not been performed experimentally, even fifty
years after the BCS theory.

During recent years the issue of the quasi-particle dynamics has
attracted considerable interest in the field of high-temperature
superconductors.\cite{Basov05} While a general agreement exists
that no well-defined superconducting gap opens, there remain
questions of the residual absorption and the narrow quasi-particle
mode which evolves with an extremely small scattering rate at low
temperatures.\cite{Pimenov99, Turner03} It therefore seems worth to revisit
the issue of quasi-particle response in a conventional
superconductor. Here we give the first report on the temperature
and frequency dependence of the complex conductivity of aluminum
in a very wide energy range ($3\cdot 10^{-4}<
\hbar\omega/2\Delta(0)=h f/2\Delta(0)<0.3$ and $T/T_c > 0.5$); we investigate the
dynamics of the quasi-particles and the influence of the mean free
path.

\begin{table}[b]
\caption{Characteristic parameters of the aluminum films. $d$ is
the film thickness, $p$ is the controlled oxygen pressure during evaporation, $Z$ is
the sample impedance at $T=300$~K, $\rho$ is the room temperature resistivity,
$\ell$ refers to the mean free path; the low-temperature skin depth for $f=40$~GHz is denoted by $\delta$; $T_c$ is the critical
temperature, $\xi(0)$ is the effective coherence length, and $\lambda(0)$ is the effective penetration depth.
\label{table}}
\begin{ruledtabular}
\begin{tabular}{lccccccccc}
 &$d$ & $p$ & $Z$ & $\rho$ & $\ell$ & $\delta$ & $T_c$ & $\xi(0)$ & $\lambda(0)$\\
& (nm) &  ($\mu$Torr) & ($\Omega$) & ($\mu\Omega$cm) & (nm)& ($\mu$m) & (K) & (nm)& (nm)\\
\hline\\
A & 40 &$1.2$  &1.65 & 32 & 5.0   & 15&1.70 &61 & 200\\
B & 30 &$20$   &1.25 & 43  & 3.7 & 17&1.75&53 & 228\\
C & 50 & $30$    &2.90  & 87& 1.8 & 24&1.90 &35 & 310\\
\end{tabular}
\end{ruledtabular}
\end{table}

\section{Experiment}
\begin{figure}
\includegraphics[width=70mm]{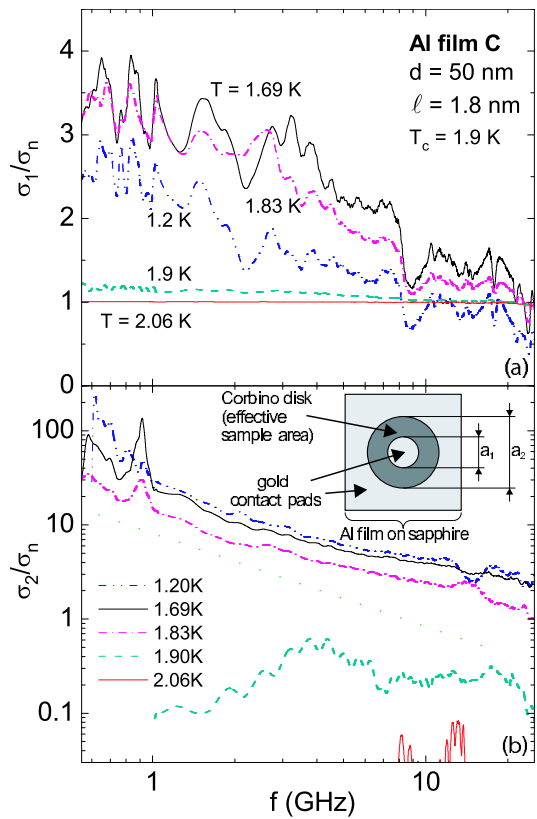}
\caption{\label{fig:sigomega}(color online) Frequency dependence of the complex conductivity of the aluminum film C ($T_c = 1.9$~K), normalized to the metallic conductivity
$\sigma_n = \sigma_1(T=2.1~\textrm{K})$. (a) Real part of the conductivity; note that $\sigma_1$ does not vary monotonously with temperature, but goes through a maximum around $T\approx 1.69$~K. (b) The imaginary part is plotted on a double-logarithmic scale in order to demonstrate the $\sigma_2\propto 1/f$ behavior (green dots). The inset is a sketch of the Corbino arrangement: $a_1 = 0.8$~mm, $a_2 = 1.75$~mm.}
\end{figure}

Aluminum films of different thickness (30 to 50~nm) were thermally
evaporated (rate 1~nm/s) onto sapphire substrates. The film
thickness was determined by a quartz microbalance, atomic force
microscopy, and ellipsometry. In order to vary the mean free path,
the films were prepared at different partial pressures of oxygen,
as listed in Tab.~\ref{table}. For technical reasons, the films
were exposed to air, resulting in an Al$_2$O$_3$ layer of
approximately 2~nm, before electrical contacts were made by
evaporating 100~nm thick gold pads which adapt the geometry of the
coaxial connector used for the high-frequency measurements (cf.\
inset of Fig.~\ref{fig:sigomega}). The critical temperature $T_c$
was determined from the resistance between the contacts; the width
of the superconducting transition is typically 10~mK.

\begin{figure}
\includegraphics[width=70mm]{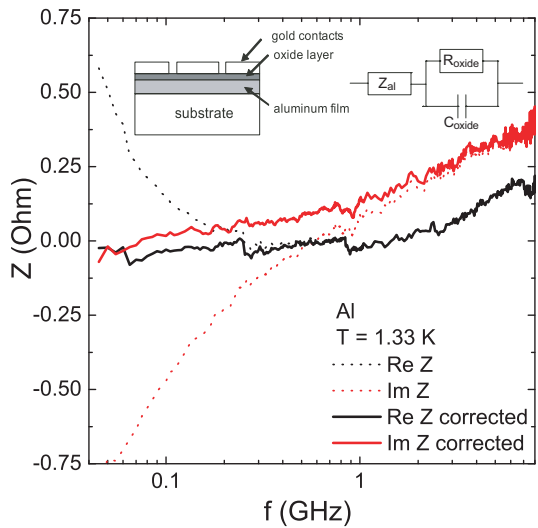}
\caption{\label{fig:imp}(color online) Real and imaginary parts of
the frequency-dependent impedance $Z$. The dashed lines indicate
the uncorrected impedance of a superconducting aluminum film. The
solid lines show the impedance after the correction for the oxide
layer between aluminum film and gold contacts. This correction is
performed assuming the lumped circuit shown in the inset: $Z_{al}$
is the impedance of the aluminum film under study whereas $R_{\rm
oxide}$ and $C_{\rm oxide}$ represent the resistive and capacity
contribution of the oxide layer.}
\end{figure}

The microwave conductivity was measured with a Corbino
spectrometer based on a HP8510C vector network analyzer in the
range from 45 MHz to 40 GHz.\cite{Scheffler05}) The data analysis
-- from complex reflection coefficient via sample impedance to
complex conductivity  -- is done for each frequency separately and
requires no additional assumptions except the sample being thin
compared to the skin depth $\delta$ (which in the complete
frequency and temperature range exceeds $10~\mu$m). For cryogenic
measurements ($T>1.1$~K) a full three-standards calibration was
performed. Bulk aluminum and a teflon disk were used as short and
open end, respectively. NiCr films (25~nm thickness, 80\%\ Ni and
20\%\ Cr) serve as a frequency-independent load ($Z\approx
7~\Omega$). The calibration by the short is most crucial; checks
in the normal state of aluminum (where the impedance is frequency
independent) and data analysis with different calibrations rule
out artifacts. The thin oxide layer between the Al film and the Au
contacts acts as a capacitor with a noticeable but correctable effect below 5~GHz; its influence
vanishes for higher frequencies. The influence of the oxide layer
is temperature independent and the additional impedance can be
subtracted from the sample impedance. Fig.~\ref{fig:imp}
demonstrates the effect of the additional capacitor in the
superconducting state. While at higher frequency the curves become
identical, the corrections are appreciable at low frequencies.

The mean free path $\ell$ of our films is evaluated from the dc
resistivity $\rho$ in the normal state: $\ell = v_F m / (n e^2
\rho)$ with the carrier concentration $n= 6.45 \cdot 10^{22}~{\rm
cm}^{-3}$ and mass $m=1.4 m_e$ ($m_e$ is the free electron mass).
\cite{Ashcroft76} The coherence length is related to the
transition temperature: $\xi_0=0.18\hbar v_F/(k_BT_c)$, where
$v_F=2.03\cdot 10^{8}$~cm/s is the Fermi velocity. The values are
listed in Tab.~\ref{table}. Because the samples are in the
so-called dirty local limit,\cite{Tinkham96,DresselGruner02} both
the effective coherence length $\xi(0)=0.855\sqrt{\xi_0\ell}$ and
penetration depth $\lambda(0)=\lambda_L\sqrt{\xi_0/(1.33 \ell)}$
are larger than the London penetration depth of bulk aluminum
$\lambda_L=15$~nm. For all films the condition  $\ell < \xi(0) <
\lambda(0)$ is easily fulfilled. Our results are in good agreement
with previous investigations of granular Al films.\cite{Cohen68}

\begin{figure}
\includegraphics[width=85mm]{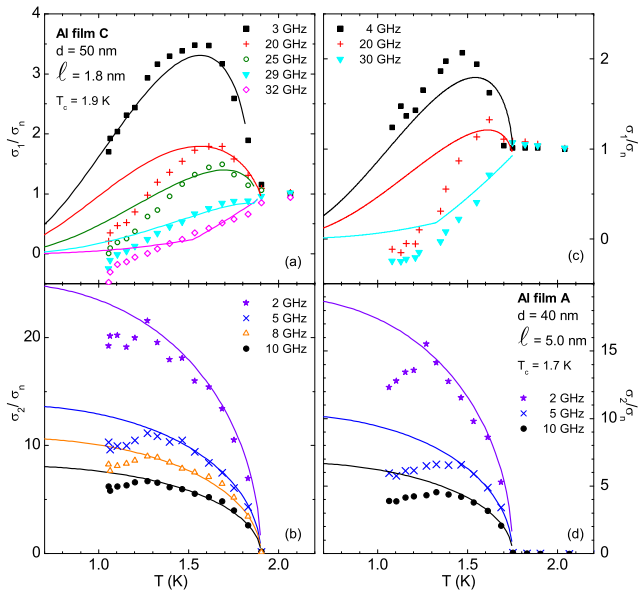}
\caption{\label{fig:sigT}(color online) Temperature dependence of
the (a and c) real and (b and d) imaginary parts of the
conductivity of aluminum normalized to the normal state
conductivity $\sigma_n$. The frames (a) and (b) show the data
measured on the Sample C at various frequencies as indicated. The
solid lines are calculated by the BCS  using $\ell/(\pi\xi)
=0.03$. The frequencies are chosen in a way to demonstrate the
effects most clearly. The results of Sample A are displayed in
panels (c) and (d). The respective calculations are performed with
$\ell/(\pi\xi) =0.1$.}
\end{figure}

\section{Results}
\subsection{Conductivity}
\begin{figure}
\includegraphics[width=60mm]{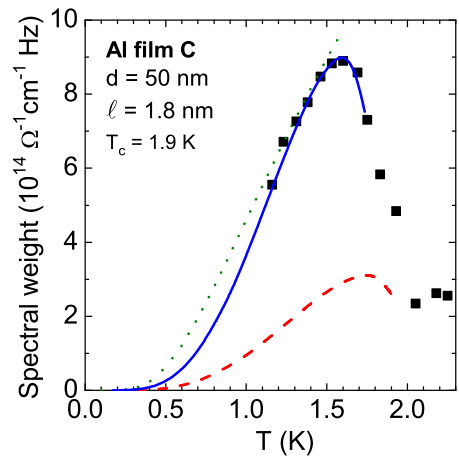}
\caption{\label{fig:spectralweight}(color online) Temperature
dependence of the spectral weight $I^s(T)$ for the Sample C
(squares) evaluated according to
Eq.~(\protect\ref{eq:spectralweight}). The dashed red line
represents the spectral weight obtained by integrating the
conductivity calculated by the BCS model (cf.\
Fig.~\ref{fig:BCS3}) up to 40 GHz. If we scale the theoretical
curve in temperature and spectral weight to the maximum of the
data (shown by the solid blue line), the description of the
experimental values is excellent. The increase below $T_c$ is due
to the enhanced density of states as the superconducting gap
opens. After a maximum is reached around 1.6~K, the spectral
weight collapses because fewer quasi-particles can be excited
across the gap when $T\rightarrow 0$. The low-temperature behavior
can be approximated by a exponential law (dotted green line).}
\end{figure}

The frequency-dependent real and imaginary parts of the normalized
conductivity are plotted in Fig.~\ref{fig:sigomega} for the film
C, as an example. For the metallic state conductivity, we have
chosen  $\sigma_1 (T=2.1~{\rm K})$ independent of frequency.
$\sigma_1/\sigma_n$ increases rapidly as the temperature is
reduced below $T_c=1.9$~K, it goes through a maximum around 1.69~K
and then drops again. Theory \cite{Tinkham96,DresselGruner02}
predicts a $\left(\sigma_1/\sigma_n\right)_{\rm max}\propto
\log\left\{2\Delta(0)/\hbar\omega\right\}$ frequency behavior of
the maximum, and experimentally we observe this with the roughly
linear curve for $T=1.69$~K in Fig.~\ref{fig:sigomega}a. An even
better description is obtained by a Drude response with a
non-monotonous temperature dependence of the spectral weight as
will be discussed below (inset of Fig.~\ref{fig:scatter} and
Fig.~\ref{fig:spectralweight}).

The imaginary part $\sigma_2$ describes the response of the Cooper
pairs to the electric field. For small frequencies
\begin{equation}
\frac{\sigma_2(\omega,T)}{\sigma_n} = \frac{\pi \Delta(T)}{\hbar\omega} \tanh\left\{\frac{\Delta(T)}{2 k_B T}\right\}
\label{eq:sigma2}
\end{equation}
in good agreement with our findings plotted in
Fig.~\ref{fig:sigomega}b, in particular the linear contribution
can be seen by comparing with the dotted line in that figure. In
Fig.~\ref{fig:sigT}b and d, $\sigma_2/\sigma_n$ is plotted for
various frequencies as a function of temperature. Not too close to
$T_c$, when $\Delta >2k_BT$, the temperature dependence of
$\sigma_2$ basically follows the opening of the superconducting
gap $\Delta(T)$, according to Eq.~(\ref{eq:sigma2}). For all
frequencies, the behavior can be fitted by the BCS prediction,
assuming $\ell/(\pi\xi) = 0.1$ for Film A and 0.03 for Film C, in
fair agreement with the values calculated in Tab.~\ref{table}.

\begin{figure}
\includegraphics[width=70mm]{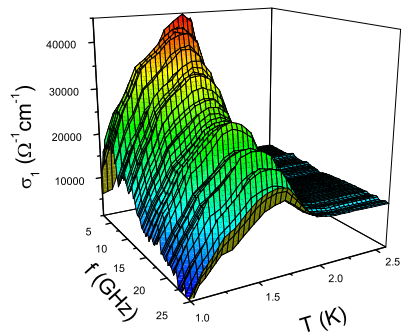}
\caption{\label{fig:3D}(color online) Real part of the
conductivity of aluminum film C ($T_c = 1.9$~K) as a function of
frequency and temperature. The coherence peak decreases in height
as the frequency increases. For comparison with
Fig.~\ref{fig:BCS}, it should be noted that the superconducting
energy gap $2\Delta(0)\approx 140$~GHz, which implies that we
focus on the very low frequency part
[$\hbar\omega/2\Delta(0)<0.2$]. }
\end{figure}
\begin{figure}
\includegraphics[width=50mm]{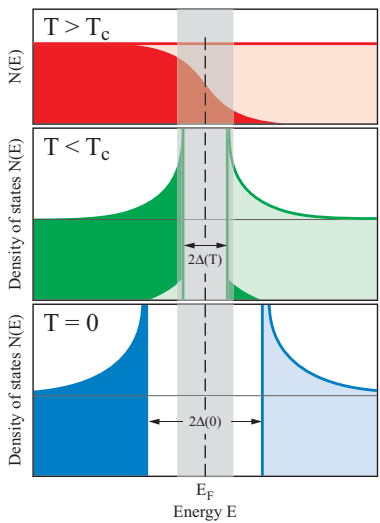}
\caption{\label{fig:DOS}(color online) Density of electronic
states around the Fermi energy $E_F$. In the normal state
($T>T_c$), the density of states is basically constant; the states
are occupied according to the Fermi distribution (dark filled
area). For $T<T_c$ the gap opens, and the density of states
exhibits a singularity. At finite temperatures there are still
states occupied above $E_F$ and state empty below; these are the
ones which contribute to the quasi-particle conductivity. For
$T=0$ the gap is complete and no quasi-particle excitations are
possible up to $\hbar\omega=2\Delta(0)$. The grey-shaded area
depicts the range which corresponds to the finite energy
$\hbar\omega$ of our microwave radiation. It can be considered as
a smearing of the border between occupied and empty states
similar to the thermal broadening.}
\end{figure}

Figs.~\ref{fig:sigT}a and c show data (taken on Sample C and A at
different frequencies) of the temperature dependence of the real
part of the conductivity. $\sigma_1$ is governed by charge
carriers thermally excited across the gap; their density of states
diverges at the gap edge as depicted in Fig.~\ref{fig:DOS}. Right
below $T_c$, $\Delta(T)$ is small: the thermal energy but also the
photon energy $\hbar\omega$ are sufficient to break up Cooper
pairs.
The coherence factor $F({\cal E},{\cal E'})$, which
describes the quasi-particle scattering, depends on their energy
 ${\cal E}$ and ${\cal E'}$. If summed over all
{\bf k} values, it reads \cite{Tinkham96} $F(\Delta,{\cal E},{\cal
E'})=\frac{1}{2}\left(1+{\Delta^2}/{{\cal EE'}}\right)$; only for
energies close to the gap $\Delta$, this factor is appreciable:
$F\approx 1$. Hence the coherence peak is seen as a maximum in
$\sigma_1(T)$ at approximately $0.8~T_c$ in the low-frequency
limit; it becomes smaller with increasing frequency and shifts to
higher temperatures; this is clearly observed in
Fig.~\ref{fig:sigT}. Above 28~GHz [corresponding to
$0.2\cdot\Delta(0)$] the peak is completely suppressed, and for
higher frequencies $\sigma_1(T)$ monotonously drops below $T_c$.
Our findings of the frequency- and temperature-dependent
conductivity of Al are summarized in the three-dimensional
representation of Fig.~\ref{fig:3D}; we can follow the evolution
of the coherence peak over a substantial part of the
temperature-frequency space.

\subsection{Spectral Weight}
Based on general electrodynamics,\cite{DresselGruner02} the spectral weight
\begin{equation}
I=\int \sigma_1(\omega)\,{\rm d}\omega= \frac{\pi n
e^2}{2m} \label{eq:spectralweight}
\end{equation}
is a measure of the charge carrier density $n$. In
Fig.~\ref{fig:spectralweight} we plot the spectral weight $I^s$
for the superconducting state by integrating the experimental data
of $\sigma_1$ between 1 and 40~GHz. \cite{remark1} We see that the
effective carrier concentration increases below the transition
temperature, passes through a maximum around 1.6~K and vanishes
for $T\rightarrow 0$ in an exponential fashion. This behavior is
perfectly explained by the BCS theory (dashed and full curve in
Fig.~\ref{fig:spectralweight}). Here Fig.~\ref{fig:DOS} may serve
as an illustration: As the temperature drops below $T_c$ the
superconducting gap opens gradually. Due to the finite temperature
$T$, there are still states occupied above the gap and empty
states below. These are subject to single particle excitation at
arbitrarily small energies, which leads to the quasi-particle
contribution in the optical conductivity. The divergence in the
density of states causes the increase of that part of spectral Of
course, there is a second contribution to $I^s(T)$ given by the
electrons excited across the superconducting gap $2\Delta(T)$ (but
not accessed in our experiment).

\begin{figure}
\includegraphics[width=70mm]{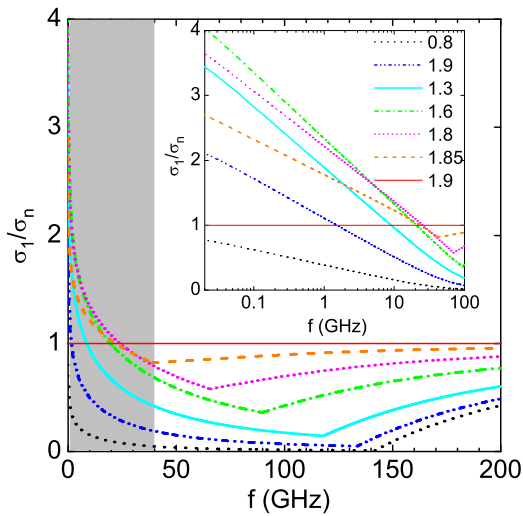}
\caption{\label{fig:BCS3}(color online) Frequency-dependent
conductivity of aluminum at different temperatures calculated
according to the BCS theory,\cite{Zimmermann91} formulated by
Mattis and Bardeen.\protect\cite{Mattis58} For the critical
temperature we assumed $T_c=1.9$~K, the superconducting energy gap
$2\Delta(0)/h = 140$~GHz. The inset, with the clear correspondence
to our experimental data in Fig.~\ref{fig:sigomega}a, illustrates
the rapid growth of the zero-frequency quasi-particle peak as the
temperature drops below $T_c$; for $T< 1.6$~K it becomes narrower
and eventually vanishes as $T\rightarrow 0$. The frequency range
covered by our experiments is indicated by the grey-shaded area.}
\end{figure}

The corresponding conductivity is plotted in Fig.~\ref{fig:BCS3}.
The calculations are done according to the BCS
model,\cite{Zimmermann91} using the formulas of Mattis and
Bardeen.\cite{Mattis58} To closely simulate the behavior of
aluminum, we assume a transition temperature $T_c=1.9$~K; the
superconducting energy gap is given by $2\Delta(0)/h = 140$~GHz.
As the temperature drops slightly below $T_c$, a narrow
zero-frequency mode builds up which grows continuously until it
reaches a maximum around $T=1.6$~K. As the temperature is reduced
further, two effects cause the spectral weight (as summed over the
experimentally accessible frequency range, the shaded region in
Fig.~\ref{fig:BCS3}) of the single-particles to decrease: the
energy gap becomes larger and the Fermi distribution sharpens.
Consequently, the number of quasi-particles available for
transport, and accordingly $I^s(T)$, decreases and finally
vanishes for $T\rightarrow 0$. As demonstated in
Fig.~\ref{fig:spectralweight} the theoretical curve (solid line)
nicely describes the experimental findings (squares).

We can describe the drop of spectral weight by the exponential
freezing out of the normal carriers; as demonstrated by the dotted
line in Fig.~\ref{fig:spectralweight}. Since the total spectral
weight has to be conserved \cite{DresselGruner02} this drop has to
be recovered. The Tinkham-Glover-Ferrell sum rule
\cite{Glover56,Ferrell58} states that the spectral weight
\begin{equation}
A=I^n-I^s=\int_{+0}^{\infty}\left[ \sigma_1^n-\sigma_1^s\right]
{\rm d}\omega
\end{equation}
missing in the superconducting state below the gap is transferred
to the $\delta$-peak at $\omega=0$.

\subsection{Scattering Rate}
\begin{figure}
\includegraphics[width=60mm]{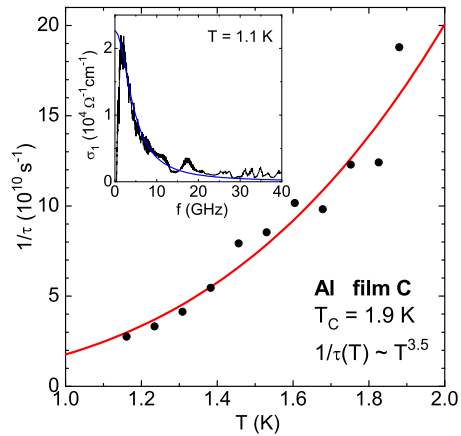}
\caption{\label{fig:scatter} The temperature-dependent scattering
rate for the aluminum Film C. The solid line corresponds to a fit
$1/\tau=A+B\cdot T^{\alpha}$ with $A=0$, $B=(17.6 \pm 3.3)\cdot
10^{10}$ and $\alpha=3.5 \pm 0.3$. The inset shows an example of
a conductivity spectrum ($T=1.116$~K) fitted by the Drude model
Eq.~(\ref{eq:Drude}).}
\end{figure}

By now we have assumed that the real part of the conductivity is
solely determined by the density of states and coherence factor;
i.e.\ the quasi-particle concentration $n$ increases and
eventually vanishes according to the temperature dependence of the
spectral weight $I(T)$ defined in Eq.~(\ref{eq:spectralweight}):
$n(T)=(2/\pi) I(T)\, m /e^2$. Also the spectral behavior of the
conductivity $\sigma_1(\omega)$ is given by the density of states
$N(E)$. In the course of exploring the microwave properties of
high-temperature superconductors, a debate arose whether
information on the quasi-particle scattering can be
obtained.\cite{Bonn92,Klein94c,Bonn94} In fact, we can fit the
real part of the optical conductivity $\sigma_1 (\omega)$ by the
Drude formula \cite{DresselGruner02}
\begin{equation}
\sigma_1(\omega)=\frac{ne^2 \tau}{m}\frac{1}{1+(\omega \tau)^2}
\label{eq:Drude}
\end{equation}
quite well at least for the lowest temperatures (inset of
Fig.~\ref{fig:scatter}) and may extract a width which is related
to the scattering rate $1/\tau$ within this model. The results for
Film C are plotted in Fig.~\ref{fig:scatter} for different
temperatures.  The temperature-dependent scattering rate can be
fitted by a power-law $1/\tau(T) \propto T^{\alpha}$ with the
power $\alpha=3.5 \pm 0.3$. Similar observations for the case of
niobium have been reported previously.\cite{Klein94,Klein94c} It
should be pointed out, however, that this approach assumes a
constant density of states, which might change with temperature
according to the two-fluid model or similar models, but exhibits
no energy dependence in the vicinity of the Fermi energy $E_F$.

\subsection{Influence of the Mean Free Path}
\begin{figure}
\includegraphics[width=50mm]{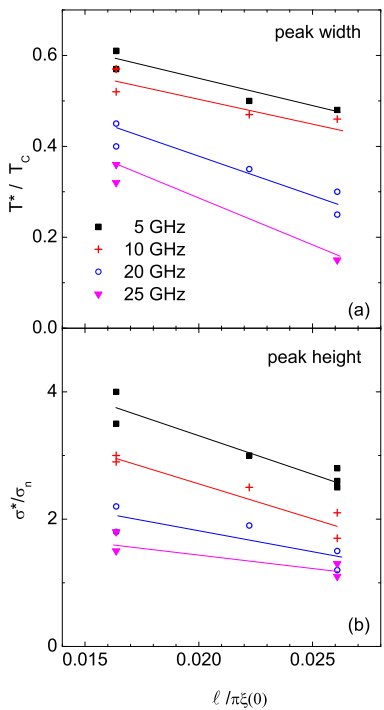}
\caption{\label{fig:peak}(color online) (a) Width $(T_c -T_n)/T_c$
and (b) height $(\sigma_1)_{\rm max}/\sigma_n$ of the conductivity
coherence peak for aluminum films with different mean free path
$\ell$, normalized to the effective coherence length $\xi(0)$. The
data are taken at various frequencies from 5 to 25 GHz; the lines
correspond to a linear interpolation. The multiple data points
indicate independent measurements of different samples.}
\end{figure}

In order to get more information on the scattering effects, we
prepared films with different defect concentration. The mean free
path of the normal state carriers strongly influences the
superconducting behavior and the dynamics of the quasi-particles.
Accordingly the complex conductivity appreciably varies for
samples with different $\ell$. Let $\sigma^*=(\sigma_1)_{\rm max}$
be the maximum of the coherence peak and $T^*$ the width, defined
as the temperature difference $T^*=T_c - T_n$ for which
$\sigma_1(T=T_n)=\sigma_n$ again.   In Fig.~\ref{fig:peak}a the peak
width is plotted as a function of mean free path as obtained from
Al films grown with different oxygen pressure. Since also the
critical temperature $T_c$ and the energy gap $\Delta$ depend on
the thickness and quality of the films (Tab.~\ref{table}), $\ell$
is normalized to the effective coherence length. The width of the
coherence peak $T^*$ is a measure of how fast the energy gap
opens. But it also tells how significant the coherence factor $F$
is in the scattering process: with increasing $\ell/(\pi\xi)$ the
finite-frequency conductivity $\sigma_1(\omega)$ decreases more
rapidly.
The height of the coherence peak $\sigma^*/\sigma_n$ is displayed
in Fig.~\ref{fig:peak}b, for the different films. When
$\ell/(\pi\xi)$ gets larger, $\sigma^*$ decreases; or with other
words, with increasing sample quality the coherence peak becomes
smaller. The effect is more dramatic for lower frequencies. The
ratio of mean free path and coherence length basically indicates
whether the scattering takes place within the range in which the
phase coherence of the superconducting wavefunction is observed;
only then coherence effects matter. Eventually we approach the
clean-limit superconductor, for which changes in the ac
conductivity below $T_c$ become increasingly difficult to
detect.\cite{Turner03} In a clean-limit superconductor, less
spectral weight  is transferred from the gap region to the
$\delta$-peak.\cite{Tinkham96} According to $\lambda=c/\sqrt{8A}$,
the reduced spectral weight of the $\delta$-peak is in agreement
with an increase of the penetration depth, as listed in
Tab.~\ref{table}.

\section{Conclusions and Outlook}
In conclusion, by performing high-precision microwave experiments
in a large spectral range down to low temperatures, we were able
to map the frequency and temperature dependence of the
conductivity coherence peak of aluminum with intentionally reduced
mean free path. The influence of the quasi-particle scattering
summarized in the coherence effects are elucidated by varying the
mean free path of the carriers. For large $\ell$, aluminum
approaches a clean-limit superconductor and the coherence peak
vanishes. The Tinkham-Glover-Ferrell sum rule is obeyed. Our
experiments provide a unique possibility to investigate within one
material system how scattering and the variation of the mean free
path influences the superconducting properties.

Our experiments show that microwave spectroscopy allows for a
detailed observation of coherence and scattering properties in
superconductors. While our present work focussed on the
well-known, conventional superconductor aluminum, future studies
can be devoted to unconventional superconductors.

\section{Acknowledgments}
We thank M. Dumm, J. Pflaum, and E. Ritz for help and discussions.
The theoretical curves were created by the FORTRAN program of E.H.
Brandt (Stuttgart). The work was supported by the DFG.


\begin{thebibliography}{99}
\bibitem{remark1}Extrapolating the data to zero-frequency and expanding the integration below 1~GHz does not change the findings qualitatively. The error in absolute value remains below 5\%. As depicted in Fig.~\protect\ref{fig:BCS}, the upper cutoff frequency falls right in the range where the quasi-particle contribution to the optical conductivity is already considerably reduced: $\sigma_1^s < \sigma_1^n$.
\end{thebibliography}
\end{document}